\documentclass[manuscript]{emulateapj}
\usepackage{amsmath}
\usepackage{natbib}
\usepackage{hyperref}
\usepackage{color}

\newcommand{\mail}{liliray@pa.msu.edu}
\newcommand{\flux}{\,erg\,cm$^{-2}$\,s$^{-1}$}
\newcommand{\lum}{\,erg\,s$^{-1}$}

\newcommand{\cm}{\,cm$^{-2}$}
\newcommand{\nh}{$N_\mathrm{H}$}

\shorttitle{\textit{Chandra} and \textit{Swift} X-ray observations of the X-ray pulsar SMC X-2 during the outburst of 2015}
\shortauthors{Li et al.}

\begin{document}
\title{\textit{Chandra} and \textit{Swift} X-ray observations of the X-ray pulsar SMC X-2 during the outburst of 2015}
\author{
K. L. Li\altaffilmark{1}, C.-P Hu\altaffilmark{2}, L. C. C. Lin \altaffilmark{3}, and Albert K. H. Kong\altaffilmark{4}
}

\altaffiltext{1}{Department of Physics and Astronomy, Michigan State University, East Lansing, MI 48824, USA; \href{mailto:\mail}{\mail}}
\altaffiltext{2}{Department of Physics, University of Hong Kong, Pokfulam Road, Hong Kong, China}
\altaffiltext{3}{Institute of Astronomy and Astrophysics, Academia Sinica, Taiwan}
\altaffiltext{4}{Institute of Astronomy and Department of Physics, National Tsing Hua University, Taiwan}

\begin{abstract}
We report the \textit{Chandra}/HRC-S and \textit{Swift}/XRT observations for the 2015 outburst of the high-mass X-ray binary (HMXB) pulsar in the Small Magellanic Cloud, SMC X-2. While previous studies suggested that either an O star or a Be star in the field is the high-mass companion of SMC X-2, our \textit{Chandra}/HRC-S image unambiguously confirms the O-type star as the true optical counterpart. 
Using the \textit{Swift}/XRT observations, we extracted accurate orbital parameters of the pulsar binary through a time of arrivals (TOAs) analysis. In addition, there were two X-ray dips near the inferior conjunction, which are possibly caused by eclipses or an ionized high-density shadow wind near the companion's surface. Finally, we propose that an outflow driven by the radiation pressure from day $\sim$10 played an important role in the X-ray/optical evolution of the outburst. 
\end{abstract}

\section{Introduction}
SMC X-2 is a high-mass X-ray binary pulsar in the Small Magellanic Cloud (SMC), discovered during an outburst in 1977 by the \textit{SAS~3} X-ray observatory in October of the year. The outburst X-ray luminosity (2--11~keV) was $1.0\times10^{38}$\lum\ with a distance of $d=71$~kpc (i.e., $7.4\times10^{37}$\lum\ with $d=62.1$~kpc; \citealt{2014ApJ...780...59G}), which was the second brightest X-ray source in the SMC. The transient was revisited by \textit{SAS~3} two months after the outburst, but was undetected with a 3$\sigma$ upper limit of $\sim10^{37}$\lum. Since then, SMC X-2 was missed by the SMC X-ray survey with the \textit{Einstein} Observatory in 1979--1980 \citep{1981ApJ...243..736S}, but re-detected by ROSAT (0.15--2.4~keV) at $L_X=2.57\times10^{37}\,(d/65~\mathrm{kpc})^2$\lum\ in April 1992 \citep{1996AA...312..919K}. In the early 2000, another X-ray outburst was detected by the RXTE/ASM with a peak luminosity of $>10^{38}$\lum\ and the subsequent RXTE/PCA observations discovered an X-ray coherent signal at a period of 2.37~s to first show SMC X-2 as a transient X-ray pulsar \citep{2001ApJ...548L..41C}. In addition to the RXTE data, the 2.37~s X-ray pulsations were also detected by ASCA in April 2000 \citep{2001PASJ...53..227Y}, reconfirming the pulsar nature of SMC X-2. Both RXTE and ASCA observations reveal hard spectra of the SMC X-2 with best-fit photon indices ranging from $\Gamma\approx0.7-1.0$. Additionally, an emission excess at 6.3~keV was found in the ASCA spectrum, which was suggested to be the fluorescent line from neutral or low-ionization iron. 

Owing to the limited ASCA localization of SMC X-2, the optical counterpart was not conclusively identified. Either an O star and a Be star, which are respectively at the north and south direction of SMC X-2 with a separation of 2.5$\arcsec$, is the true high-mass companion \citep{2006AJ....132..971S}. With 4 years of semi-continuous OGLE III data, a periodicity of $P=18.62\pm0.02$~d was detected for the northern O star while a relatively stable lightcurve is found for the southern Be star \citep{2011MNRAS.412..391S}. The periodicity was thought to be associated the orbital period of SMC X-2. The orbital period of SMC X-2 was later found to be $P=18.38\pm0.02$~d through the RXTE spin-frequency history \citep{2011MNRAS.416.1556T} that is close to the OGLE period, however, with a 0.24~d difference\footnote{Both the uncertainties of the spin period measured with OGLE and RXTE presented in \cite{2011MNRAS.412..391S} and \cite{2011MNRAS.416.1556T} are only of statistical origin. Possibly but not certainly, the periods could be consistent with each other if the systematic errors were considered. }. 

Until recently, the new X-ray outburst of 2015 detected by MAXI (a.k.a., MAXI~J0051-736; \citealt{2015ATel.8088....1N}), and later confirmed by \textit{Swift} \citep{2015ATel.8091....1K} and INTEGRAL \citep{2015ATel.8207....1F}, has provided an opportunity to distinguish the candidate optical counterparts. Followed up by an intensive \textit{Swift}/XRT ToO monitoring campaign, a $\sim800$~ks stacked \textit{Swift}/XRT image (PC mode) suggested that the northern O star is more likely to be the real counterpart \citep{2015ATel.8091....1K}, though the result is heavily based on the astrometric mapping solution between the \textit{Swift}/XRT and UVOT, and a higher spatial-resolution X-ray data is required to confirm the result. In this paper, we present a high-resolution \textit{Chandra}/HRC-S image for an unambiguous optical counterpart identification. In addition, \textit{Swift}/XRT observations in Windowed Timing (WT) mode are used to trace the 2.37~s spin-period evolution as well as to perform a phase-resolved spectral analysis during the first month of the outburst. 

\section{\textit{Chandra} observation}
\begin{figure}[h]
\centering
\includegraphics[width=85mm]{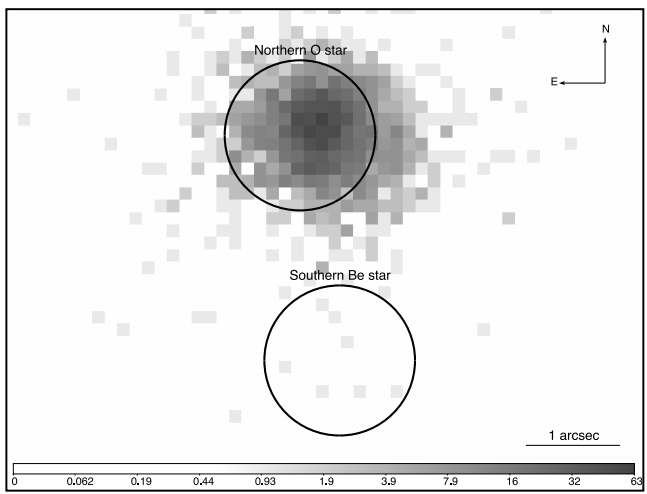}
\caption{The \textit{Chandra}/HRC-S image taken on Nov 5 2015 shows the accurate X-ray position of SMC X-2, which unambiguously indicates the northern O star as the high-mass companion of the pulsar binary. The black two circles indicates the northern and sourthern stars with the same radius of 0.8$\arcsec$ to demonstrate the 90\% confidence absolute astrometric accuracy of \textit{Chandra}. 
}
\label{fig:chandra}
\end{figure}

We requested a 2.9~ks \textit{Chandra} Director's Discretionary Time (DDT) observation taken in HRC-S timing model on 2015 November 5, which is about 43 days since the MAXI discovery (i.e., MJD 57288.6). A bright X-ray source was detected at the position of $\alpha\mathrm{(J2000)}=00^\mathrm{h}54^\mathrm{m}33\fs421$, $\delta\mathrm{(J2000)}=-73\arcdeg 41\arcmin 00\farcs99$ (in \textit{Chandra} astrometic frame), which is consistent with the previous SMC X-2 position from ASCA and \textit{Swift}/XRT. Since SMC X-2 is the only bright X-ray source in the HRC-S field-of-view, no further astrometric correction can be done on the image. We therefore adopted the \textit{Chandra} absolute astrometric accuracy, which is 0.8$\arcsec$ at 90\% confidence (Belinda Wilkes, private communication), as the positional accuracy of the source. Comparing with the northern and southern OGLE candidate counterparts, the X-ray source is aligned with the northern star strongly suggesting the brighter O-type variable (i.e., orbital period: 18.62~d) as the true high-mass companion (Figure \ref{fig:chandra}). 

\section{OGLE optical detection}
\begin{figure*}
\centering
\includegraphics[width=160mm]{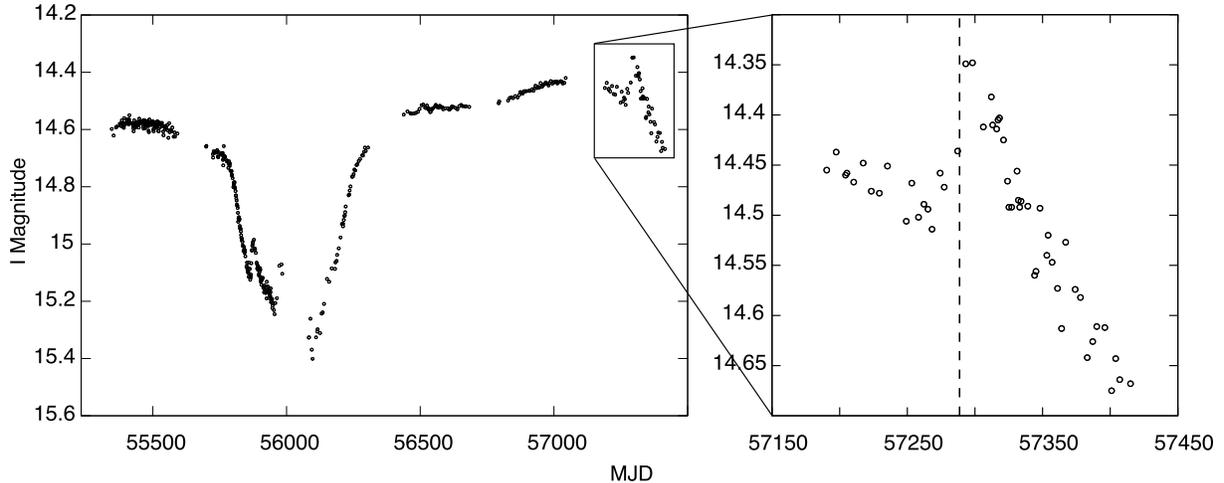}
\caption{The left panel shows OGLE I-band lightcurve from 2010 to 2015 (all the magnitude uncertainties are 0.003~mag, which is too fine to show in the plot). The right panel shows the 2015 SMC X-2 outburst with a dashed line indicating the MAXI discovery date. }
\label{fig:ogle}
\end{figure*}

Using the public data from the X-Ray variables OGLE Monitoring (XROM) system \citep{2008AcA....58..187U} of the O-type companion, the 2015 X-ray outburst of SMC X-2 is clearly seen in $I$-band (Figure \ref{fig:ogle}). The optical outburst was first seen on MJD 57274 (about 15 days before the MAXI discovery at MJD 57288.6). It then raised from $m_I=14.458\pm0.003$~mag to a peak of $m_I=14.348\pm0.003$~mag at MJD 57298.7 (about 10 days after the MAXI discovery), which is brighter by $\Delta m_I=0.166\pm0.004$~mag than the brightness measured a month ago. After that, the $I$-band flux dropped at a linear rate of $\dot{m_I}=0.0027\pm0.0001$~mag day$^{-1}$. 

\section{\textit{Swift}/XRT observations}
In this work, we used the \textit{Swift}/XRT observations in WT mode taken from 2015 September 24 to 2015 December 14 (OBIDs: 340730001--132 and 81771001--002) downloaded from the \texttt{HEASARC} data server. As SMC X-2 was piled up in the Photon Counting (PC ) mode data and the timing resolution of the PC mode data is insufficient to study the spin-period of SMC X-2, we abandoned the first-two \textit{Swift}/XRT PC mode snap shot for simplicity. Lightcurves and spectra in the energy range of 0.3--10~keV were extracted by \texttt{xrtgrblc} of \texttt{HEAsoft} version 6.17, with updated \texttt{CALDB} files (date of release: 2015-07-21). To deal with the one-dimensional WT observations with changing count rates, the X-ray events were extracted using count-rate dependent rectangle extraction regions, i.e., $35\arcsec\times15\arcsec$ for count rates between 0 and 1~cts/s, $71\arcsec\times15\arcsec$ for 1--5~cts/s, and $118\arcsec\times15\arcsec$ for 5--10~cts/s, with a corresponding pair of background regions extended along the both ends of the source region for optimizing the S/N ratios (see the \texttt{xrtgrblc} manual on the official \textit{Swift} web page for details). 
Finally, we applied a barycentric correction to the data using \texttt{barycorr} (DE200). 

\subsection{X-ray Spectral Evolution}
\label{sec:spec}
As stated in \cite{2015ATel.8091....1K}, the XRT spectra can be in general described by an absorbed power-law with a hard photon index (i.e., $\Gamma=0.61\pm1.5$ for the PC mode data; \citealt{2015ATel.8091....1K}). 
Instead of fitting a simple power-law, \cite{2016MNRAS.tmpL...3P} improved the spectral fitting by using a high energy exponential cutoff power-law model (i.e., \texttt{cutoffpl} in \texttt{XSPEC}) with an additional thermal component (i.e., a kT$\sim$0.1~keV blackbody or a kT$\sim$1~keV \texttt{apec} thermal plasma), which fitted well with the XMM-\textit{Newton} data. Additionally, the authors favour the blackbody model (instead of the plasma model), which offers a better physical interpretation of the observed X-ray lines. In this work, we adopted the \texttt{cutoffpl+bbody} in \texttt{XSPEC} to preform individual spectral fits for the \textit{Swift}/XRT observations. As the X-ray emission lines \citep{2016MNRAS.tmpL...3P} are too weak to significantly affect the overall fitting results, we did not include them in this analysis. For the X-ray absorption, only the dominant SMC absorption with an abundance of $Z=0.2~Z_\sun$ \citep{1990ApJS...74...93R} was considered. In addition, we fixed the column density and the cut-off energy to the best-fit values in \cite{2016MNRAS.tmpL...3P} (i.e., \nh\,$=1.8\times10^{21}$\cm\ and $E_\mathrm{cutoff}=6.9$~keV; \citealt{2015ATel.8091....1K}) to simplify the model. We also rejected low-count spectra (i.e., number of data bins smaller than 40 after binning with factors of at least 20 counts) to ensure the data quality. Finally, photons with energies below 0.6~keV were ignored to avoid any artificial low-energy spectral excess in the WT Mode spectra\footnote{\url{http://www.swift.ac.uk/analysis/xrt/digest_cal.php}}. 

\begin{figure}[h]
\centering
\includegraphics[width=85mm]{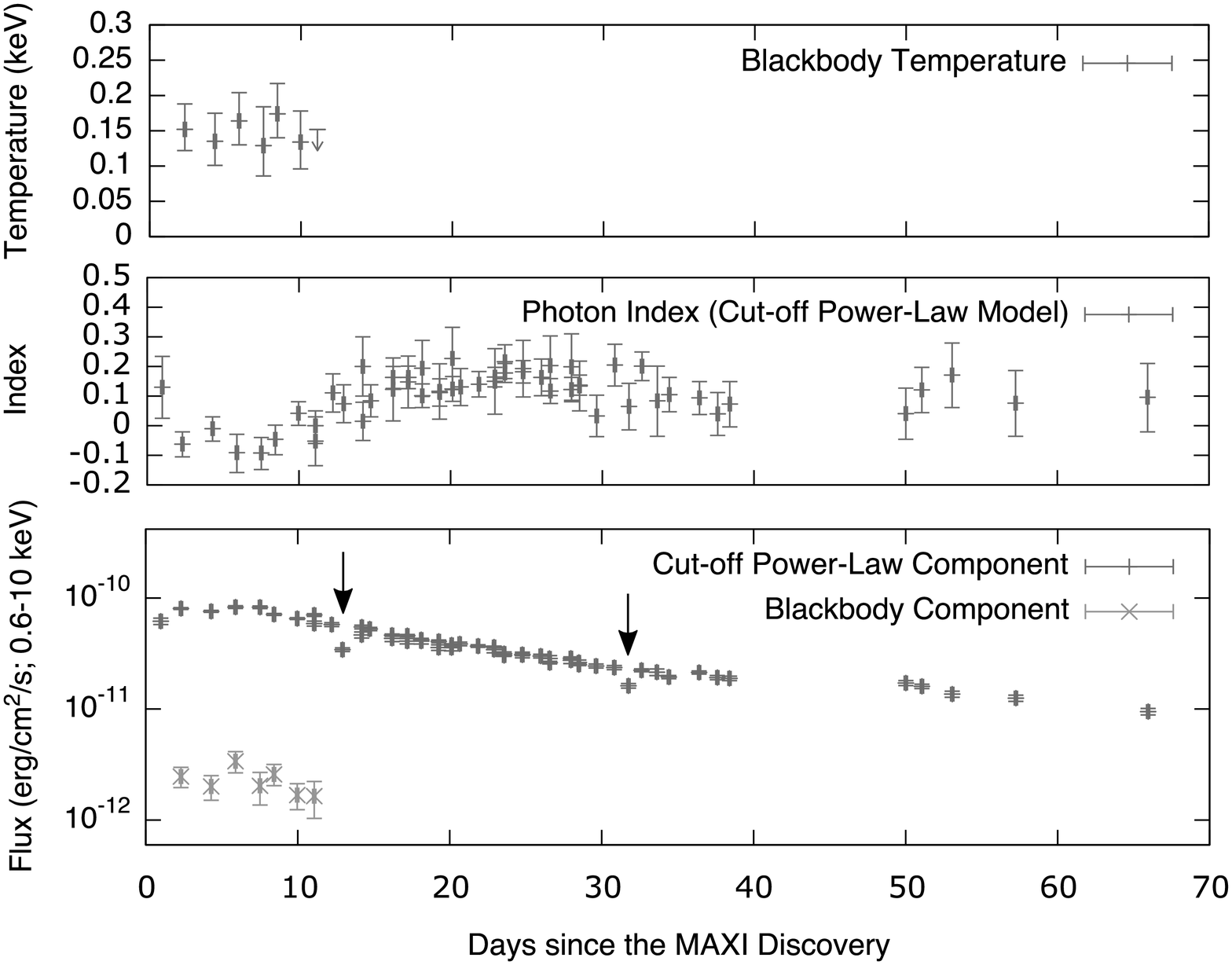}
\caption{The figures show the X-ray evolution of the SMC X-2 outburst of 2015. From the top to bottom, the panels are the best-fit blackbody temperatures, photon indices of a cut-off power-law, and the (non-)thermal X-ray fluxes. In the third panel, there are two X-ray dips indicated by two black arrows, which are probably caused by an ionized high-density shadow wind. All the uncertainties are at 90\% confidence. 
}
\label{fig:swift_curves}
\end{figure}

Except for the first observation with a low exposure time, a soft blackbody X-ray emission was clearly seen in the data taken within the first ten days. The thermal emission was last detected marginally on day 11. Since then, the thermal component was invisible (i.e., uncertainties of the thermal component go unacceptably large) and we removed the thermal component accordingly. 

Most of the fits are good with reduced chi-square values of $\chi^2_\nu=$0.9--1.2 and even the worst fit has a chi-square value of $\chi^2_\nu=1.4$ ($dof=165$). For those six fits with a thermal component, all best-fit temperatures are around $kT\approx0.15$~keV, which are roughly consistent $kT=0.135^{+0.014}_{-0.011}$~keV measured by XMM-\textit{Newton} on day 15 \citep{2016MNRAS.tmpL...3P}. For the non-thermal component, the photon index was complicatedly evolving. It first dropped from $\Gamma\approx0.1$ on day 1 to $\Gamma\lessapprox0$ around day 2--10, then raised back to a $\Gamma\approx0.2$ plateau on day 15 for about 15 days. There is a seemingly decreasing trend from the $\Gamma\approx0.2$ plateau to $\Gamma\approx0.1$ on about day 38. For the last few data, the signal-to-noises are insufficient to see significant trends (Figure \ref{fig:swift_curves}). 

The X-ray flux of SMC-X-2 was dominated by the non-thermal component during the whole outburst as we have learned from the XMM-\textit{Newton} data \citep{2016MNRAS.tmpL...3P}. The X-ray flux was raising and hanging at around $4-6\times10^{-10}$\flux\ during the first eight days and the X-ray emissions were continuously decreasing with $F_X\sim t^{-1.10\pm0.05}$ (Figure \ref{fig:swift_curves}). In addition, two X-ray dips were clearly seen at days 13.0 and 31.8, with only about 60\% of the inferred flux observed (Figure \ref{fig:swift_curves}). 


\subsection{X-ray Timing Analyses}
The \textit{Swift}/XRT observations in WT mode with a fine time resolution of 1.766~ms of the long-term monitoring are extremely useful to investigate the 2.37~s spin-period previously detected by RXTE and ASCA \citep{2001ApJ...548L..41C,2001PASJ...53..227Y}, and hence probe the orbital parameters of the binary by tracking the observed spin period. In fact, \cite{2016MNRAS.tmpL...3P} used the same set of \textit{Swift}/XRT data to compute an orbital period of $P=18.38\pm0.96$ days and a projected semi-major axis of $a \sin i=78\pm3$ lt-s through a sinusoidal fit to the changing spin period. Here we re-analyse the \textit{Swift} timing data in an alternative way, namely time of arrivals (TOAs), to examine the SMC X-2's ephemeris during the 2015 outburst. 

To obtain a completely precise ephemeris for SMC X-2, we performed a measurement of pulse time of arrival (TOAs) determined by the \textit{Swift} data and then fitted all the parameters of a timing model to these measurements. We first used H-test to measure the spin period in each observation and noted that the pulse can barely be detected in the second orbital cycle with H-values less
than 50. Moreover, the evolution of those significantly detected pulsed periods looks like a sinusoidal function owing to the orbital \textit{Doppler} effect. We then fitted the spin periods using a sinusoidal curve to obtain a set of initial parameters of orbital motion. The initial timing parameters are already good and consistent with the result derived by \citealt{2016MNRAS.tmpL...3P}, but folding the individual data set with them can yield a visible drifting of pulse phase. Therefore, we further improved the orbital parameters using the TOA analysis. 

\begin{figure}[h]
\centering
\includegraphics[width=85mm]{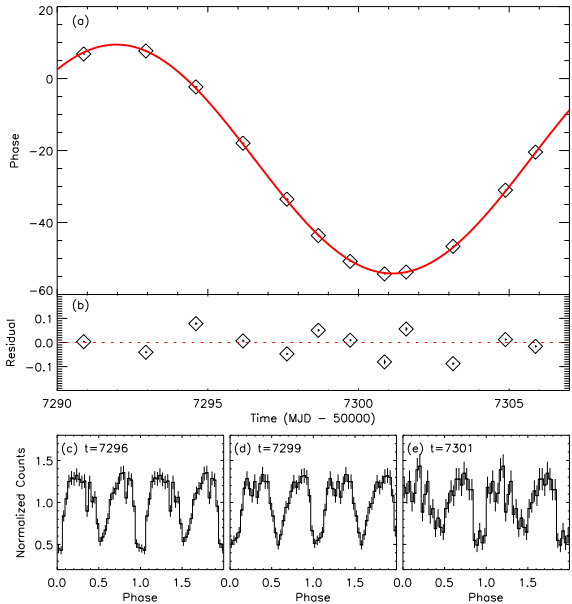}
\caption{The upper panel shows the phase-shift evolution of the measured minimum phases as a function of time and the best-fit binary model in Table \ref{tab:timing} (a) with residuals (b). Three examples of the best (c), a typical (d), and the worst (e) measurements of the minimum phase are shown in the lower panel. 
}
\label{fig:phase_res}

\end{figure}

In the TOA analyses, we folded the X-ray photons to investigate the pulse profile in each observation according to the obtained initial timing parameters. Because the pulse profile contains two flat maxima/peaks and two sharp minima/valleys, we defined the deepest minimum as the fiducial point. We then fitted the folded light curve with two sinusoidal profiles to search for possible positions of the valley points, and then described data points in the neighborhood with a Gaussian function to determine the phase and uncertainty of each minimum. The entire method is similar to \cite{2011ApJS..194...17R} to determine the timing ephemerides of \textit{Fermi} gamma-ray pulsars, and the only difference is that they used the pulsed peak in the obtained profiles to determine the pulse TOAs. We collected 12 pulse minima in the first orbital cycle to be used in the TOA analysis (Figure \ref{fig:phase_res}). The data points of the second orbital cycle are not included in the analysis because the pulse profile is barely seen. Following the iteration process described in \cite{2008ApJ...678.1316C}, we improved the timing parameters assuming an elliptical orbital model. The final result is shown in Table \ref{tab:timing} and the phase residual is shown in Figure \ref{fig:phase_res}. The small phase residuals ($\sim$0.05) further validate the derived timing solution. We found that a finite eccentricity of $e=0.019\pm0.006$ can improve the fitting, but a circular orbit can already provide a convincing result in the timing model of SMC X-2. The orbital period is determined as $18.33\pm0.17$ days and the projected semi-major axis is $75 \pm 1$ light-seconds. The orbital parameters contain considerable uncertainties because those TOAs determined in the first orbital cycle dominate the fitting to describe the timing model. Further deep monitoring of this pulsar would help us to more precisely constrain the orbital properties. 

\begin{table}[]
\centering
\caption{Rotational and orbital parameters of SMC X-2}
\begin{tabular}{l}
\hline
Freq. (pulsar frequency) = 0.421553(1) Hz \\
$a \sin i/c$ (projected semi-major axis) = 75(1) lt-s \\
$P$ (binary orbital period) = 18.33(17) day \\
$T_{\pi/2}$ (epoch of 90$^\circ$ mean longitude) = MJD 57282.8(1) \\
$e$ (orbital eccentricity) = 0.019(6) \\
$\omega$ (longitude of periastron) = 0.58(18) \\
\hline
 \label{tab:timing}
\end{tabular}
\end{table}

\subsubsection{Comparison with the \textit{Chandra} Pulse Profile}
\begin{figure}[h]
\centering
\includegraphics[width=85mm]{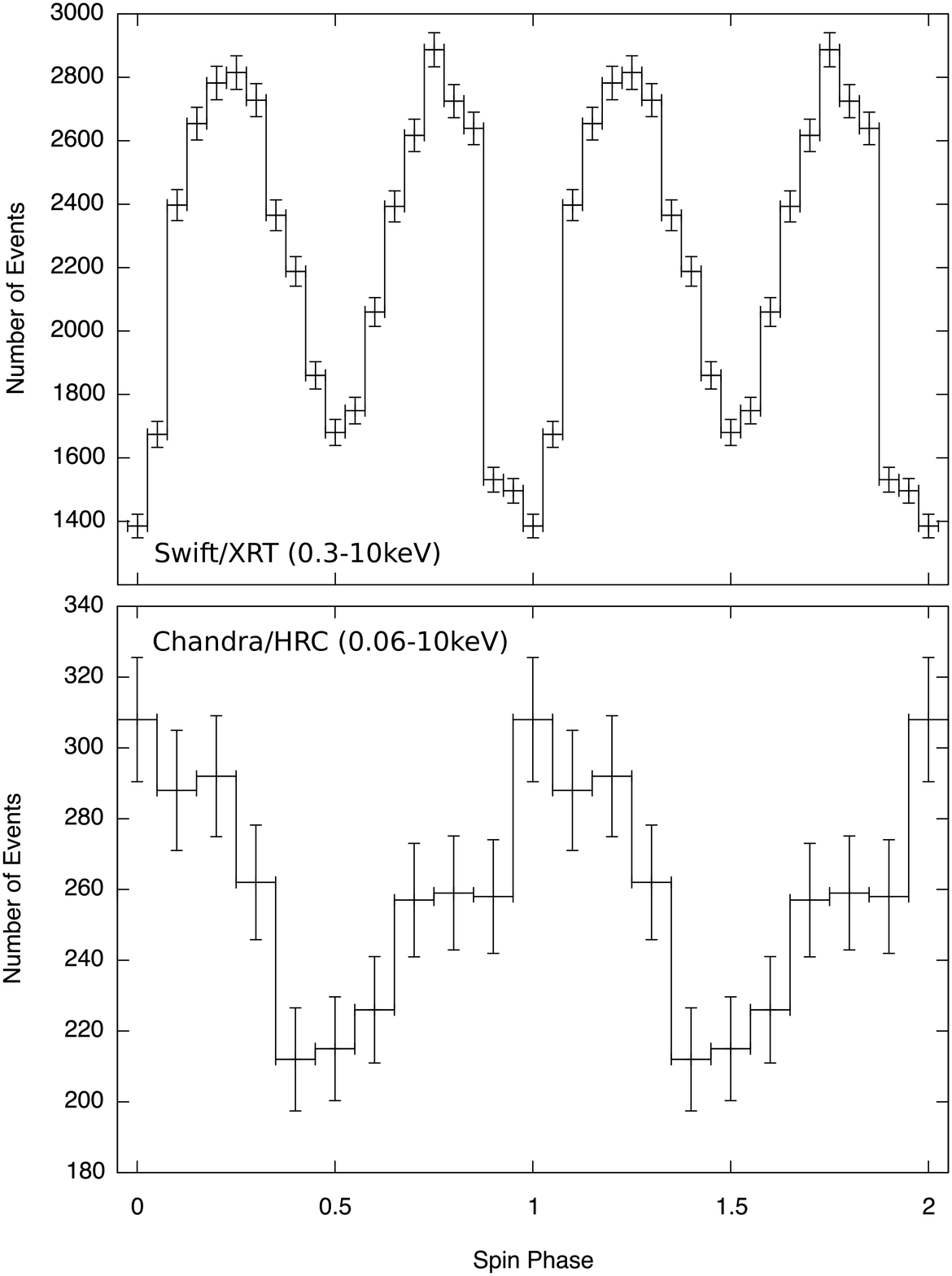}
\caption{The X-ray pulse profiles of SMC X-2 using the barycentre-corrected events extracted from, i) the five selected \textit{Swift}/XRT observations in WT mode (0.3--10~keV), and ii) the \textit{Chandra} HRC-S data (0.06--10 keV). Two cycles are shown for clarity. Caution: the \textit{Chandra} and the \textit{Swift} pulse phases are not aligned. 
}
\label{fig:profile}
\end{figure}

We analysed the HRC-S timing mode data to investigate the X-ray pulse profile in 0.06--10 keV at the later phase of the outburst. 
A signal of H-value = 28 is detected at $p=2.3723\pm0.0005$~s and corresponds to a $>4$ sigma confidence level for a single trial. Since the observing length is short (i.e., the \textit{Fourier} resolution is $2\times10^{-3}$~s) and the SMC X-2 was relatively faint (i.e., 2577 useful events in a 2.9~ks exposure), the \textit{Chandra} period is not accurate enough to reveal the orbital effect, but the best-fit period is consistent with the spin periodicity of our model (cf., Table \ref{tab:timing}) in 1$\sigma$ confidence level. 
We folded the \textit{Chandra} data and obtained a likely single-pulsed profile as shown in Figure \ref{fig:profile}.
Considering the fact that our timing model may be not applicable to the \textit{Chandra} epoch due to a possible change of the spin periodicity after the outburst, we do not stress on the phase difference between the \textit{Chandra} and \textit{Swift} profiles but only concentrate on the shape of pulse profile. 
Even we tried different periods within the \textit{Fourier} resolution to fold the \textit{Chandra} data, the profiles are all single peaked as presented in Figure \ref{fig:profile}, although the width of peaks and valleys may change. We also checked the \textit{Swift} data in the later phase and found similar single-pulsed profiles occasionally shown up, although the apparent single-pulsed profiles could be in part due to the low photon statistics. 

\begin{table}[]
\centering
\caption{Rotational-phase resolved spectra of SMC X-2}
\begin{tabular}{llll}
\hline
Phase& 0.075--0.375 &0.625--0.875 & Off-pulse \\
\hline \\
\nh\ & \multicolumn{3}{c}{$1.8\times10^{21}$\cm\ (fixed)} \\[2ex]
$E_\mathrm{cutoff}$ & \multicolumn{3}{c}{6.9~keV (fixed)} \\[2ex]
$\Gamma_\mathrm{cutoff}$& $-0.06\pm0.03$ & $-0.07\pm0.03$ & $-0.04^{+0.04}_{-0.05}$ \\[2ex]
$T_\mathrm{bb}$ (keV)& $0.13^{+0.04}_{-0.03}$ & $0.14\pm0.03$ & $0.17\pm0.02$ \\[2ex]
$F_\mathrm{cutoff}^{1}$ (0.6--10~keV)& $6.70\pm0.14$ & $7.92^{+0.17}_{-0.16}$ & $3.68\pm0.09$ \\[2ex]
$F_\mathrm{bb}^{1}$ (0.6--10~keV)& $0.05^{+0.02}_{-0.01}$ & $0.07\pm0.02$ & $0.08\pm0.01$ \\[2ex]
$\chi^2/dof$& \multicolumn{3}{c}{1376.52/1407} \\[2ex]
 \hline
 \multicolumn{4}{l}{$^{1}$ All the fluxes are in $10^{-10}$\flux. }\\
 \multicolumn{4}{l}{$^{2}$ All the stated uncertainties are at 90\% confidence. }\\
 \label{tab:spec}
\end{tabular}
\end{table}

\subsection{Phase-resolved Spectra}
We chose five observations (i.e., OBIDs: 340730002--003, 005, and 007--008) to stack a high-quality spectrum (i.e., about 43~k spectral data counts) as representative data to perform a spin-phase-resolved spectral fitting analysis. The representative data were selected based on the brightness of SMC X-2 during the observation (i.e., count-rates higher than 6 cts s$^{-1}$) and the exposure length (i.e., $>$ 1~ks). Besides achieving good photon statistics, all the selected data were taken within a week to minimise probable spectral variations among the observation as the outburst evolved. Using our timing solution (Table \ref{tab:timing}), we computed the spin-phase of the events accordingly to resolve the data in phases. The X-ray pulse profile is shown in the top panel of Figure \ref{fig:profile}. We also defined the phase intervals of $\sigma_s=0.075-0.375$ and $0.625-0.875$ to be the first and second peaks (i.e., bins with a count number larger than the average of all the bins), and the rest to be the off-pulse intervals. The absorbed \texttt{cutoffpl+bbody} model was used and the fitting results are listed in the Table \ref{tab:spec}. No obvious changes are seen in the cut-off index and the temperature among the phases, however, the two on-pulse non-thermal X-ray emissions are significantly higher than the off-pulse emission. Since the thermal fluxes in the on/off-pulse phases are very close, we conclude that the X-ray pulsations were dominantly driven by the non-thermal component. 

\section{Discussion and Conclusion}
It has been 15 years since the last outburst of SMC X-2 in 2000. For the 2015 outburst, we made use of the data from the long-term \textit{Swift}/XRT monitoring campaign as well as the \textit{Chandra}/HRC-S observations to investigate SMC X-2 at various aspects. The key findings presented in this paper are summarized and discussed in the following: 

\subsection{Identification of the O-type Companion}
Using the high-spatial resolution \textit{Chandra} data, we unambiguously confirm the northern O-type star, instead of the southern Be star, previous detected by OGLE III \citep{2006AJ....132..971S}, is the high-mass companion of the pulsar system. Besides, the OGLE detection of the 2015 outburst provides independent evidence of the association. 

\subsection{The Spin Period Change in 15 years}
By the analysis with TOAs, we greatly improved (5 times more accurate) the rotational and orbital parameters of SMC X-2 in 2015 (Table \ref{tab:timing}), comparing with the ones in \cite{2016MNRAS.tmpL...3P}. We found a $18.33\pm0.17$ days spin-period modulation caused by the orbital motion of the pulsar through \textit{Doppler} shifting ($18.38\pm0.96$ days in \citealt{2016MNRAS.tmpL...3P}), which is consistent with the value of $18.38\pm0.02$d measured by RXTE in 2000 \citep{2011MNRAS.416.1556T}. 

However, despite the spin-up trend seen in the 2000 outburst by RXTE (i.e., $\dot{p}=(-7.20\pm0.15)\times10^{-11}\,$s s$^{-1}$), the spin period measured with \textit{Swift} in 2015 is significantly larger than the spin period measured with RXTE, indicating that the pulsar was actually spinning-down at $\dot{p}=(4.9\pm0.2)\times10^{-13}\,$s s$^{-1}$ over the last 15 years. The spin-down value presented in this work is slightly lower than the one in \cite{2016MNRAS.tmpL...3P} (i.e., $\dot{p}=(6.6\pm0.2)\times10^{-13}\,$s s$^{-1}$) because of the different corrections for the orbital motion. Nevertheless, the spinning-up observed in 2000 is likely a transient spin change during the X-ray outburst as a consequence of angular momentum transferred from the accreting matter to the pulsar. Similar phenomena have been widely seen in several binary pulsars during their X-ray outbursts (e.g., KS 1947+300; \citealt{2005AstL...31...88T}). 

By comparing the \textit{Swift}/XRT and \textit{Chandra} pulse profiles, the X-ray pulsations of SMC X-2 likely evolved from a double-peaked profile to a single-peaked profile (Figure \ref{fig:profile}). 
In fact, three \textit{NuSTAR} observations taken on 2015 September 25, October 12/21 clearly showed that the profile (3--79~keV) was evolving over time \citep{Jaisawal25052016} and the most prominent change can be seen in the third observation (i.e., 15 days before the $Chandra$ data). Although this last \textit{NuSTAR} profile is still double-peaked, it is generally flatter and one of minima is significantly wider and deeper than the other one (see Figure 2 in \citealt{Jaisawal25052016} for details). From this pulse profile evolution, it would not be a big surprise to see a nearly single-peaked profile of \textit{Chandra} 15 days later as the weaker minimum was too shallow to be resolved by \textit{Chandra} in 2.9~ks. Single-peaked X-ray pulsars are not uncommon (e.g., see the HMXB pulsars 1E1145.1-6141, and GRO J1008-57; \citealt{2008AA...479..533F,2014RAA....14..565W}), but such a dramatic evolution on the pulse profile is rare. Moreover, a similar single-pulsed profile of SMC X-2 has been seen below 2~keV by ASCA in 2000 \citep{2001PASJ...53..227Y}, however, the physics behind is still unclear. 

\subsection{The X-ray Dips at the Inferior Conjunction}
Two X-ray dips were detected in the \textit{Swift}/XRT lightcurve at days 13.0 and 31.8 (\S \ref{sec:spec}). Assuming an orbital period of 18.33~d (phase zero: MJD 57282.80), the corresponding orbital phases ($\phi$) are 0.024--0.025 (duration: 1.6~ks) and 0.050--0.058 (duration: 13~ks), which are both around the inferior conjunction of the HMXB (i.e., along the line of sight: observer--companion--pulsar). 
In the following, we discuss several possible mechanisms to produce the X-ray dips. 
While \cite{2008MNRAS.388.1198M} has shown that the SMC X-2's companion has a stellar type of O9.5 III--V, we tentatively apply a well-studied star of similar type, the O7.5 III--V primary star of the binary $\gamma^2$ Velorum (i.e., $M=28.5\,M_\sun$ and $R=17\,R_\sun$; \citealt{2007MNRAS.377..415N}) to simulate the companion for easier discussions. 

\subsubsection{X-ray Eclipse}
Pulsar eclipses may occur and cause the \textit{Swift} X-ray dips, if the inclination angle $i$ is large enough (i.e., edge-on if $i=90^\circ$). 
In this scenario, the binary inclination ($i$) and the companion's size ($R_*$) can be simply constrained by the relation, $R_*\gtrsim a\,\cos i$. The critical inclination angle would be approximately $i\gtrsim58^\circ$ with assumptions of $R_*\approx17\,R_\sun$ and $a\,\sin i/c=75$~light-seconds (Table \ref{tab:timing}). 

However, after checking the lightcurve in details, we found no obvious X-ray dip detected in the observation of day 50 (the first data after the 12-day observing gap; Figure \ref{fig:swift_curves}), at which (i.e., $\phi=0.046-0.050$) an eclipse should have also been seen. Moreover, the observation before the second X-ray dip was actually taken closer to the inferior conjunction (i.e., $\phi=-0.0001-0.004$), but without showing any X-ray dip either. If the X-ray dips were really caused by eclipse, the ``X-ray shadow'' (along the line-of-sight) should be symmetric about the phase zero (but the dips only occur at $\phi > 0.004$) and should not evolve over time (but the dip disappears around day 50). 

The X-ray dips can be more symmetric by fine tuning the ephemeris within the uncertainties. For example, using the largest $P$ and $T_{\pi/2}$ allowed in Table \ref{tab:timing} (i.e., 18.50d and MJD 57282.9) would lead the X-ray dips closer to the mid-eclipse (i.e., $\phi=0.008-0.009$ and $\phi=0.024-0.032$) and almost eliminate the $\phi > 0.004$ constraint (i.e., the relevant pre-dip observation would then be in the interval from $\phi=-0.026$ to $-0.022$). However, this does not explain the absence of the dip around day 50, which is expected given the relatively small phase difference between the second dip and the day 50 observation (i.e., $\Delta\phi=0.005-0.01$ for $P=18.16-18.50$d). Also the eclipsing scenario requires a large inclination of $i\gtrsim58^\circ$ but the binary mass function suggests in the opposite way (i.e., $i\approx22^\circ$, see Section \ref{sec:bmf}). While we cannot completely rule out the eclipsing scenario in this work, it is still unfavored based on the analytical results derived from the current observations. 

\subsubsection{A Clumpy Stellar Wind}
Wind clumps are often formed in HMXBs as the wind driving mechanism (i.e., acceleration by radiation pressure) is a highly unstable one leading to irregularities in the distributions of density and velocity \citep{2002A&A...381.1015R,2013A&A...560A..32B}. High-density wind clumps are strong X-ray absorbers and they could be origins of some X-ray dip phenomena. In fact, this scenario has been proposed to explain the X-ray dips near the inferior conjunction seen in Cyg X-1 (i.e., a black hole HMXB; \citealt{2002ApJ...564..953F}). In the case of SMC X-2, the X-ray dips might be produced when a wind clump happened to enter the line-of-slight and moved away from the line-of-sight on day 50 so that no dip was observed then. Through this scenario, the X-rays are attenuated by photoelectric absorption, which should cause an energy-dependent reduction at low energies (i.e., no X-rays below 1~keV should have been seen in the SMC X-2's case). 
But no such energy-dependent reduction has been seen in the dip observations (see the second panel of Figure \ref{fig:swift_curves} for the cut-off power-law index evolution), suggesting that the clumpy wind scenario is unlikely the correct picture to explain the X-ray dips. To further confirm this, we stacked the two spectra of the dips and compared the best-fit \nh\ of it with that of the spectrum stacked using the nearest observations taken before/after the dips. The best-fit \nh\ are $2.8^{+1.8}_{-1.6}\times10^{21}$\cm\ (dip) and $2.5^{+2.2}_{-1.9}\times10^{21}$\cm\ (pre/post-dip), of which the difference is insignificant. 

\subsubsection{A Shadow Wind}
Alternatively, the absorber could be an almost fully-ionized stellar wind stalled around the X-ray-illuminated face of the companion. For an X-ray-luminous HMXB like SMC X-2, the radiative driving force of the companion's stellar wind can be suppressed by photonionization as there will be less energy states of the wind to absorb the UV photons from the star (see \citealt{1994ApJ...435..756B,2000MNRAS.311..861B} and the references therein for details). 
\cite{1994ApJ...435..756B} have performed 2D simulations for Cen X-3 and found that the suppression starts to be significant at $L_X>10^{37}$\lum\ to produce a stalled wind of a high column density (i.e., \nh\ $>10^{23}$\cm\ for Cen X-3) along the line-of-sight near eclipse egress (i.e., $\phi\gtrsim 0$; as the \textit{Coriolis} force drags the wind to cause the asymmetry), namely the \textit{shadow wind} (as the wind originated from the X-ray non-irradiated surface). Although \cite{1994ApJ...435..756B} focused only on the material contributing photoelectric absorption (i.e., that is not fully ionized with an ionization parameter, $\xi<2000$), the shadow wind could actually be highly ionized especially for a luminous X-ray system, like SMC X-2 in outburst (i.e., $L_{X,\mathrm{peak}}\gtrapprox2\times10^{38}$\lum\ with $D=62.1$~kpc; \citealt{2014ApJ...780...59G}). 
If an almost fully-ionized high-density shadow wind existed in SMC X-2 during the outburst, \textit{Compton} scattering (instead of photoelectric absorption) should have attenuated the X-ray flux to produce dips around the eclipse. 
Presumably, the pulsar should be close to the edge of the high-mass companion from our point of view at $\phi=0$ so that the shadow wind can attenuate the observed X-rays significantly. With the optically-thin \textit{Compton} scattering model (\texttt{cabs}) in \texttt{XSPEC}, we found that the X-ray flux attenuation seen in the SMC X-2's dips (i.e., dropped by $\sim40$\%) can be reproduced with a \textit{fully-ionized} wind of \nh\ $\sim6\times10^{23}$\cm, which is fairly reasonable comparing to the simulations for Cen X-3. Around day 50, the X-ray luminosity decreased to $L_X\approx4\times10^{37}$\lum, at which the geometry and/or the density of the shadow wind might have changed and hence the \textit{Compton} scattering effect was not clearly observed. 

\subsubsection{Constraint by the Binary Mass Function}
\label{sec:bmf}
From the best-fit \textit{Keplerian} parameters $a \sin i$ and $P$, we found the pulsar binary mass function to be $f_1=4\pi^2(a \sin i)^3/(GP^2)=1.37\pm0.06\,M_\sun$ and the companion mass equals to $m_2= f_1 (1+q)^2 /\sin^3 i$ (i.e., $q=m_1/m_2$ here). In the case of the pulsar mass $m_1=1.35M_\sun$ \citep{1999ApJ...512..288T} and the companion mass $m_2\approx28.5\,M_\sun$, the inferred inclination is $i\approx22^\circ$.
As mentioned, this small inclination, though not accurately measured, suggests that SMC X-2 is unlikely an eclipsing binary. 

For $i\approx22^\circ$, the pulsar will be $\sim11\,R_\sun$ away from the companion surface at $\phi=0$ (from our point of view), which is probably a bit far for the shadow wind scenario. This may suggest that $m_2\approx28.5\,M_\sun$ is overestimated, leading to an underestimation of the inclination. 
One possibility to explain the overestimation of the mass is that the companion is irradiated by the X-ray source so that the surface temperature of the companion is higher than it should be. In this case, the SMC X-2's companion mass should be smaller than the typical ones of the similar stellar types to partly cause the inconsistency. Alternatively, the inconsistency may implies that the shadow wind model fails to explain the X-ray dips. In the worse case, the dips at the inferior conjunction could just be two intrinsic flux changes (e.g., a rapid decrease in accretion), which just happened to exhibit at $\phi\sim0$. While the inclination is crucial to affect the feasibility of the shadow wind model on SMC X-2, detailed simulations will be useful to determine the critical inclination for the model and hence test the whole idea in the future. 


\subsection{The Optical and X-ray Evolutions}
By fitting the XRT data with the absorbed \texttt{cutoffpl+bbody} spectral model. We confirmed the existence of the $kT\approx0.2$~keV thermal component, which was first seen in the 30~ks XMM-\textit{Newton} observation \citep{2016MNRAS.tmpL...3P}. Based on the \textit{Swift} data, the blackbody temperature over days 2--10 is $kT=0.15\pm0.02$~keV (mean and the standard deviation of the six best-fit temperatures) with $F_X=(6.0\pm2.0)\times10^{-12}$\flux\ (0.6--10~keV; Figure \ref{fig:swift_curves}) and the blackbody component was below the detection limit of \textit{Swift} after day 11. Although it was detected again by XMM-\textit{Newton} on day 15, the temperature ($kT=0.135^{+0.014}_{-0.011}$~keV) and the flux ($F_X=3.2^{+1.1}_{-0.8}\times10^{-12}$\flux, corrected to 0.6--10~keV; \citealt{2016MNRAS.tmpL...3P}) both decreased. 

\cite{2016MNRAS.tmpL...3P} proposed that the soft thermal X-rays were from the inner edge region of the accretion disk ($R_\mathrm{in}\sim10^8$~cm), in which the primary non-thermal X-rays ($L_X\approx1.4\times10^{38}$\lum) were reprocessed into the observed blackbody X-rays \citep{2004ApJ...614..881H}. 
Our phase-resolved spectral analysis clearly showed that the thermal component did not change with the spin phase and the X-ray pulsations are mainly driven by the non-thermal X-rays from the magnetosphere, which are in a good agreement with the scenario. In addition to the reprocessing model, we further propose a possible outflow from the pulsar (started on about day 10) that weaken the accretion to explain the long-term evolution of the 2015 SMC X-2 outburst. Using the SMC distance of $D=62.1$~kpc \citep{2014ApJ...780...59G}, the \textit{Swift} X-ray luminosities in the first 10 days (those with a thermal emission detected) are all $L_X\gtrapprox2\times10^{38}$\lum, which are larger than the \textit{Eddington} limit, $L_\mathrm{Edd}=1.7\times10^{38}$\lum\ for a neutron star with the canonical neutron star mass of $M=1.35M_\sun$ \citep{1999ApJ...512..288T}. An outflow driven by the radiation pressure was likely formed to resist further accretion flow and lower the primary non-thermal X-ray flux (started from day 8--11). In fact, the cut-off power-law index started evolving from the same period, (i.e., from $\Gamma\lesssim0$ to $\Gamma\approx0.2$), as a possible consequence of the weaker accretion. Obviously, the reprocessed thermal flux decreased as the primary non-thermal flux decreased. In optical, the $I$-band lights reached maximum also on day 10 (i.e., MJD 57298.2) and began to decrease linearly at $\dot{m_I}=0.0027$~mag day$^{-1}$. Assuming the $I$-band lights are the disk emissions and the decreasing flux indicating a temperature drop on the disk (as the accretion rate was decreasing), the reprocessing region should have been moving outward as the inner edge of the cooler disk were expanding. This could reduce the temperature of the reprocessed thermal X-ray component as we have seen in the \textit{Swift}/XRT and the XMM-\textit{Newton} data. 

\begin{acknowledgements}

We thank the anonymous referee for constructive comments that significantly improved the paper. 
Support for this work was provided by the National Aeronautics and Space Administration through \textit{Chandra} Award Number DD5-16078X issued by the \textit{Chandra} X-ray Observatory Center, which is operated by the Smithsonian Astrophysical Observatory for and on behalf of the National Aeronautics Space Administration under contract NAS8-03060. 
CPH was supported by a GRF grant of Hong Kong Government under 17300215. 
AKHK was supported by the grant DD5-16078X from the grant 103-2628-M-007-003-MY3 from the Ministry of Science and Technology of Taiwan. 
KLL would like to thank the director of the \textit{Chandra} X-ray Center, Belinda Wilkes, for granting our DDT request. 
\end{acknowledgements}

\bibliography{smcx2}

\end{document}